Hidden Structural Variants in ALD NbN Superconducting Trilayers Revealed by Atomistic Analysis

*Prachi Garg[1], Danqing Wang[2], Hong X. Tang[2], Baishakhi Mazumder[1]\**

[1] Department of Materials Design and Innovation, University at Buffalo-SUNY, Buffalo, NY, USA

[2] Department of Electrical and Computer Engineering, Yale University, New Haven, CT, USA

\* Corresponding author

Funding: Air Force Office of Sponsored Research under Grant no. 162023-22608 and FA9550-23-1-0688, Army Research Office under Grant No. W911NF-24-2-0240, Office Naval Research under Grant No. N00014-23-1-2021

Abstract
Microscopic inhomogeneity within superconducting films is a critical bottleneck hindering the performance and scalability of quantum circuits. All-nitride Josephson Junctions (JJs) have attracted substantial attention for their potential to provide enhanced coherence times and enable higher temperature operation. However, their performance is often limited by local variations caused by polymorphism, impurities, and interface quality. This work diagnoses atomic-scale limitations preventing superconducting NbN/AlN/NbN JJs from reaching their full potential. Electrical measurements reveal suppressed critical current density and soft onset of quasiparticle current. However, inverse proportionality between resistance and junction area confirms homogenous barrier thickness. This isolates structural and chemical variations in electrodes and barrier as the source of performance limitation. The observed characteristics are attributed to complex materials problems: NbN polymorphism, phase coexistence, and oxygen impurities. Using advanced microscopy and machine learning integrated approach, nanoscale inclusions of ε-$Nb_2N_2$ are found to coexist within dominant δ-NbN electrodes. DC performance of JJs may be affected by these defects, leading to unresolved supercurrent and soft transition to normal state. By identifying specific atomic scale defects, tracing its origin to initial film nucleation, and linking to its detrimental electrical signature, this work establishes a material-to-device correlation and provides targeted strategy for phase engineering towards reproducible, high coherence and scalable quantum devices.

Introduction
Quantum processors rely on superconducting qubits as one of the most advanced and scalable platforms for quantum information science. Unlike natural atomic systems, superconducting qubits are lithographically defined artificial atoms whose energy spectra can be engineered through circuit design, offering tunability, reproducibility, and integration into multiqubit processors.[1] However, the performance of these devices is critically limited by the quality of their constituent Josephson Junctions (JJs), where microscopic defects, interfacial disorder, and electronic inhomogeneity can directly degrade coherence times and reproducibility.[1–3] While lithographic defects and surface oxidation are now well understood sources of two-level system (TLS) loss, the stochastic formation of parasitic polymorphs at buried interfaces remains an unresolved source of qubit decoherence. Hence, the development of scalable superconducting electronics requires the ability to engineer ultrathin, defect controlled tunnel junctions with stable and reproducible properties.[4–8] Among the various architectures, all nitride trilayer systems, particularly NbN/AlN/NbN, have gained significant attention due to their compatibility with high-frequency operation, large superconducting gap energies, and robustness against oxidation.[9–12] These properties make them promising candidates for next generation quantum devices, especially in environments requiring high thermal resilience.

Despite these advantages, achieving high quality NbN based trilayers remains challenging.[13] One major difficulty lies in the phase control of niobium nitride thin films, which exhibit complex polymorphism (e.g., δ-NbN, β-$Nb_2N_2$, γ-$Nb_4N_3$, ε-$Nb_2N_2$) depending on deposition conditions.[14–18] The presence of multiple phases can potentially alter the superconducting and tunneling behavior of JJs.[19,20] Additionally, the ultrathin barrier layer (typically ~2 nm AlN) requires atomically sharp and chemically stable interfaces to suppress leakage currents and maintain coherent tunneling. To achieve the precise conformational control required for these devices, plasma enhanced Atomic Layer Deposition (ALD) has emerged as a superior processing route over traditional sputtering techniques. Recent advances in nitride ALD have unlocked high quality superconducting films with exceptional uniformity on 3D topographies. ALD, known for its angstrom scale control and uniform growth, has the ability to address these fabrication challenges.[21–24] However, despite this processing development, validating the local atomic scale phase purity within these conformal films is highly demanding for resolving the structural origin of decoherence.

Consequently, current literature has largely focused on electrical performance or morphological analysis alone, with very few studies correlating structure property relationships across all three dimensions at atomic scale resolution. In particular, the coexistence of multiple NbN related phases and their spatial distribution within the trilayer stack is underexplored.[25,26] This leaves a critical gap in understanding how the presence of transient phases, chemical inhomogeneities and structural defects affect the superconducting behavior of ALD fabricated junctions.[23,27–29] In contrast to prior ALD NbN studies that focused primarily on electrical performance or conformal film quality, our approach uniquely combines chemical, structural, and spatial analysis to uncover coexisting NbN derived phases buried within a functioning JJ. While the existence of multiple NbN polymorphs in ALD films, like β-$Nb_2N$, has been previously noted, the atomistic evidence for distinct transient phase, ε-$Nb_2N_2$ bonding environments, their depth localization, and their correlation with junction transport behavior are shown here.

In this study, a transferable metrology framework is established to analyze the phase coexistence defects, utilizing superconducting ALD grown NbN/AlN/NbN JJs for ALDmons qubits.[30] By integrating Atom Probe Tomography (APT), Scanning Transmission Electron Microscopy (STEM) and unsupervised machine learning, we decouple the complex structural variations hidden at the buried interfaces. Specifically, we identified stochastic formation of parasitic hexagonal ε-$Nb_2N_2$ polymorph in cubic δ-NbN dominated electrodes. Correlating these defects with current-voltage (I-V) characteristics reveals how these nanoscale features impact quasiparticle tunneling behavior. While prior reports have examined ALD NbN morphology and superconductivity separately, our study integrates bond level chemical mapping with device level electrical transport property, directly linking atomic phase coexistence to quantum performance. Consequently, this analytical toolkit provides an improved approach for confirming polymorph purity, offering a targeted strategy for phase engineering towards reproducible, high coherence quantum circuits.

Results and Discussion

The NbN/AlN/NbN tunnel junctions were grown using ALD, resulting superconducting qubits were fabricated by a flip chip bonding technique following the detailed procedure reported in previous work.[30] DC transport measurement of the film is represented in the I-V curve for an 8 µm junction in **Figure 1a**. It exhibits the non-linear behavior typical of a superconductor-insulator-superconductor (SIS) tunnel junction. A key observation is the absence of a supercurrent, which can be attributed to the extremely low critical current density (Jc) (1.9pA/um$^2$) making the junction highly susceptible to environmental noise. The supercurrent is, however, detectable in devices fabricated from wafers with thinner AlN barriers.[30] Additionally, within the subgap region (bias ≈ -1 mV to +1 mV), current is strongly suppressed, indicative of effective insulating barrier. Complementary chemical information on the trilayer structure was obtained from APT analysis. Atom map shown in Figure 1c resolves distinct and sharp AlN barrier layer confined between NbN electrodes, highlighting the conformality achieved by ALD. The presence of a clear inflection point near 4.5 mV in IV curve confirms that δ-NbN, with its characteristic superconducting gap, remains the dominant phase in the electrodes. This is also seen from the 2D NbN concentration ratio maps of the (d) top and (e) bottom electrodes, which reveals near-stoichiometric NbN compositions. Moreover, the room temperature resistance (R) measured across multiple junctions with areas ranging from 1.8 µm$^2$ to 64 µm$^2$ (Figure 1b) demonstrates an inverse proportionality with junction area (A). A fit to this relationship yields a constant resistance area (R.A) product of 558.5 ± 9.7 MΩ·µm$^2$. This result indicates consistent barrier thickness across the wafer.

However, despite the successful structural and geometric benchmark, the representative IV curve exhibits the absence of supercurrent and a soft, gradual onset of quasiparticle current between 3 mV and 4.5 mV, before transitioning to a linear characteristic at higher voltages (normal-state resistance, $R_n$). This discrepancy between bulk structural integrity and observed electrical performance isolates the issue to sub-stoichiometric and crystallographic inhomogeneities within the wafer. Consequently, identifying these atomic scale defects within electrodes and barrier is a crucial step in improving device reproducibility.[31]

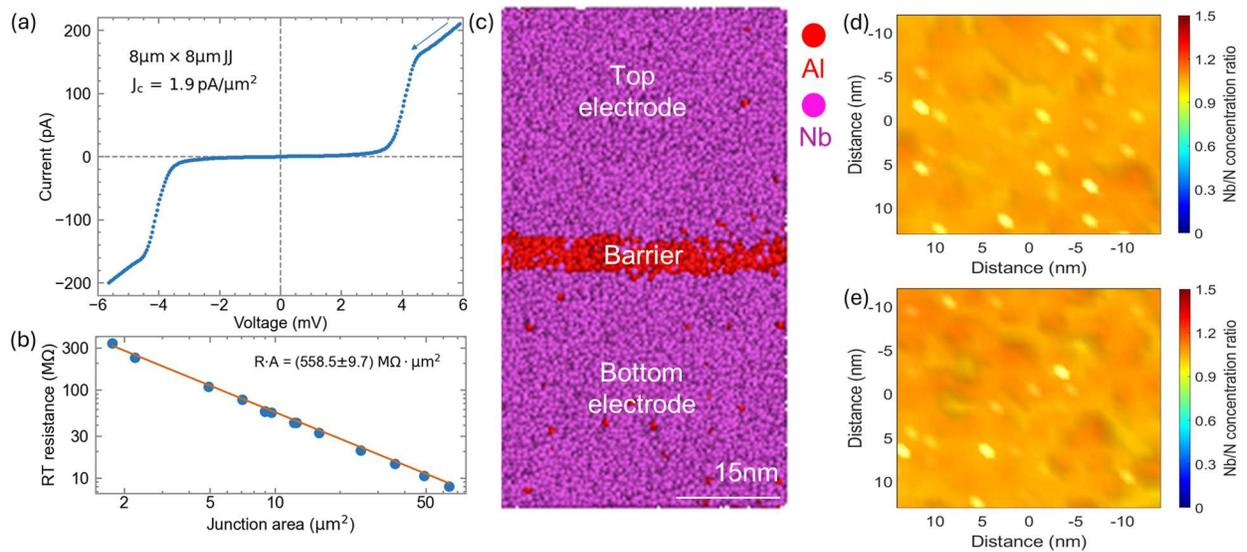

Figure 1: (a) Representative I-V curve of an ALD-fabricated NbN/AlN/NbN Josephson junction measured at 4K, with critical current density ($J_c$) of 1.9 pA/um$^2$. (b) Junction resistance dependent on area, characterized at room temperature. (c) Atomic map displaying a clear Al rich barrier layer sandwiched between Nb rich top and bottom electrodes. (d,e) 2D Nb/N concentration ratio maps from the (d) top and (e) bottom NbN electrodes, demonstrating overall stoichiometry.

A detailed elemental distribution (**Figure 2**) across the trilayer junction reveals potential microscopic issues within the barrier itself. The 3D plot visualizes the spatial distribution of oxygen and aluminum in trilayer. The 3D concentration plot in Figure 2a shows localized oxygen concentration within the AlN barrier region, while the Al distribution in Figure 2b confirms sharp barrier interfaces. The observed apparent Al rich stoichiometry in the AlN barrier layer in Figure 2c-e is a well known N undercounting artifact in wide bandgap nitrides. These are caused due to preferential field evaporation, neutral $N_2$ species, and trajectory aberrations, artificially reducing the detected nitrogen signal.[32] In contrast, the nearly stoichiometric recovery of the adjacent metallic NbN layers confirms the integrity of the analysis, indicating the Al rich ratio is a measurement bias. Although the uniform R·A product also confirms the barrier's macroscopic integrity, even low concentration of oxygen impurity at interfaces can create localized states or TLS, which are known sources of decoherence in superconducting qubits[3,33] Additionally, impurities like Al diffusing from the substrate (minor amounts observed in Figure 2c-e) or residual oxygen in the chamber (evidenced by oxygen in the barrier) may act chemically, stabilizing these transient hexagonal phases, preventing the immediate formation of pure δ-NbN. Oxygen impurities, in particular, are known to significantly affect the phase formation and superconducting properties of niobium based films.[34] These impurities may not dominate the transport characteristics observed here but represent an independent and important channel of performance degradation that must be minimized through improved ALD purity and chamber control.

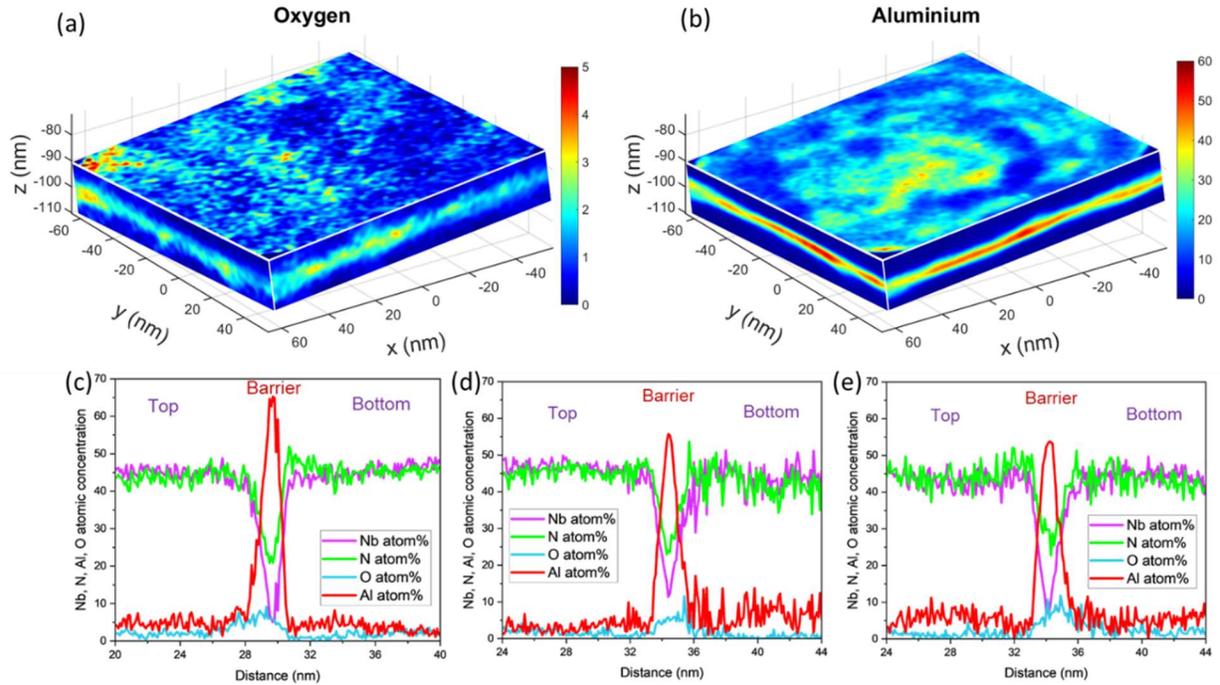

Figure 2: 3D APT maps of (a) oxygen distribution showing increased oxygen concentration localized at the AlN barrier region. (b) Aluminum distribution highlights the clear separation of the barrier from the NbN electrodes. Note that these 3D maps are projections, and features near the top or bottom electrodes may represent localized variations within the buried AlN barrier, as confirmed by the 1D concentration profiles (c-e) which clearly show oxygen enrichment confined to the barrier region.

While barrier uniformity is confirmed and oxygen represents a known decoherence channel, the specific electrical transport deviation suggests microstructural inhomogeneity within the superconducting electrodes themselves. To investigate this, TEM based clustering analysis was employed to resolve local structural variations shown in **Figure 3**. While overall film uniformity is apparent (Figure 3a), localized analysis using inverse FFT filtering (Figure 3b) and k-means clustering (Figure 3c) identifies statistically distinct regions based on lattice periodicities. The resulting d-spacings distributions (Figure 3d) identify three reproducible clusters. Crucially, two clusters exhibit median d-spacings (0.159 nm and 0.144 nm) corresponding closely to theoretical lattice spacing values for the (220) plane of δ-NbN and the (110) plane of ε-$Nb_2N_2$, respectively, validating the robustness of this classification.[35,36] The cubic δ-NbN structure's (220) plane and hexagonal ε-$Nb_2N_2$ structure's (110) plane is overlayed on the inverse FFT image (as shown in Figure 5b) of bottom electrode to confirm their coexistence. A third cluster is indicative of a mixed phase. These statistical groupings were consistent with the structural differences between the δ-NbN and ε-$Nb_2N_2$ polymorphs, which are illustrated schematically in Figure 3e. This provides direct structural evidence for the coexistence of these two polymorphs at the nanoscale within the electrodes.

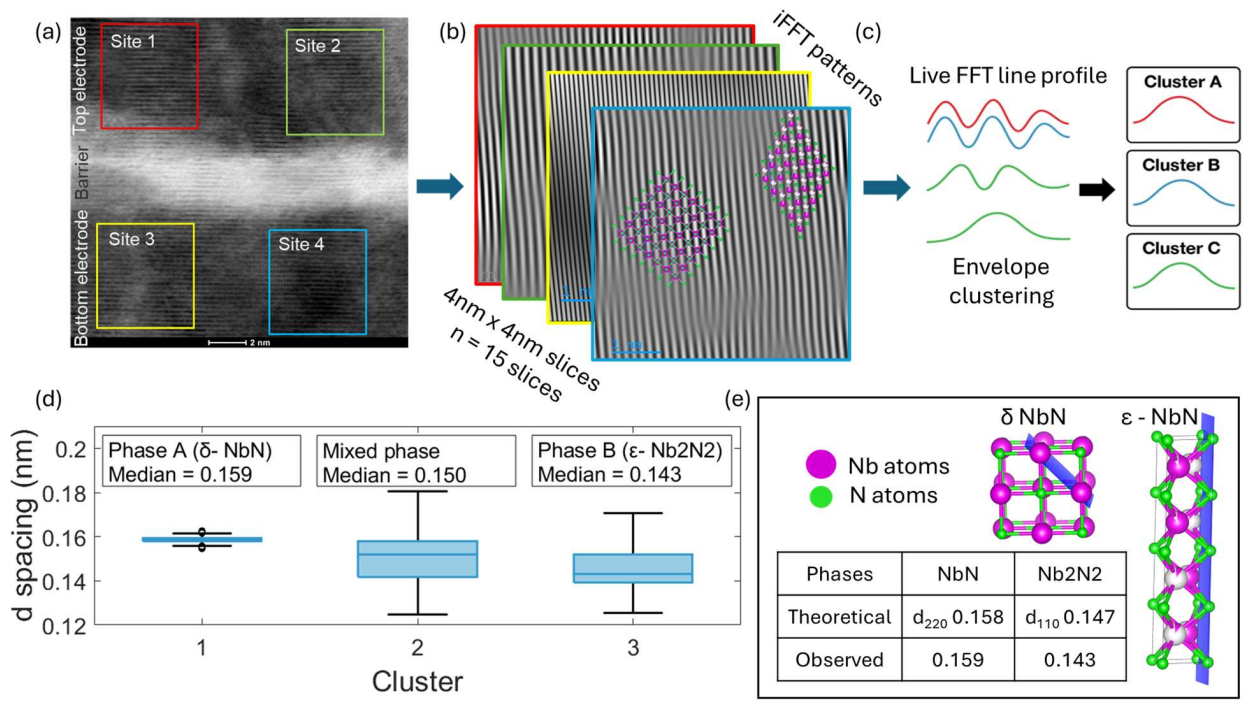

Figure 3: Workflow and results of TEM based clustering analysis of NbN/AlN/NbN trilayers. (a) TEM image showing four sites (representative) selected for local structural analysis. (b) Inverse FFT filtering was applied to isolate lattice fringes, where two phase overlays were seen. (c) Representation of envelope clustering of the line profiles that enabled grouping of regions with similar lattice periodicities into distinct clusters. (d) Boxplots of extracted d spacings reveal three clusters corresponding to Phase A (δ-NbN), Phase B (ε-$Nb_2N_2$), and a mixed region, with mean values and associated uncertainties indicated. (e) Structural schematics of δ-NbN and ε-$Nb_2N_2$ are shown for comparison, highlighting differences in lattice periodicity along the blue shaded planes at (220) and (110) for δ-NbN and ε-$Nb_2N_2$, respectively. The table at the bottom shows d spacing values calculated from theory and observed at respective orientations of δ-NbN and ε-$Nb_2N_2$.

Further corroborating evidence comes from APT detector pair separation analysis, probing local bonding environment (**Figure 4**). The workflow depicts ion evaporation and double hit detection (Figure 4a). Spatially resolved concentration maps highlight different regions showing local variation in δ-NbN (Figure 4b) and ε-$Nb_2N_2$ (Figure 4c) concentration across the APT needle. While, δ-NbN rich regions are widely distributed across the electrodes, localized domains enriched in ε-$Nb_2N_2$ are observed, often complementing sites with reduced δ-NbN concentration. Filtering for specific double hit events enabled the separation of δ-NbN - δ-NbN and ε-$Nb_2N_2$ - ε-$Nb_2N_2$ pairs, which serve as substitution for their stoichiometric bonding environments. Quantitative analysis of detector pair separations (Figure 4d) yields statistically distinct median distances of 2.77 Å for δ-NbN and 2.35 Å for ε-$Nb_2N_2$, ($p = 0.042$, $z = -1.73$), consistent with the different bonding environment and shorter Nb-Nb interatomic distances expected in hexagonal ε-phase compared to the cubic δ-phase structure.[14,19,37] The presence of these nanoscale ε-$Nb_2N_2$ inclusions provides a direct explanation for the observed electrical characteristics. The ε-$Nb_2N_2$ phase possesses distinct superconducting properties.[14,20,38,39] This nanoscale inhomogeneity is expected to locally perturb the performance of quantum devices. Furthermore, the phase boundaries between δ-NbN and ε-$Nb_2N_2$ represent structural defects known to broaden the superconducting density of states or unwanted in-gap states.[40–42] These states provide a pathway for trap assisted quasiparticle tunneling.[3]

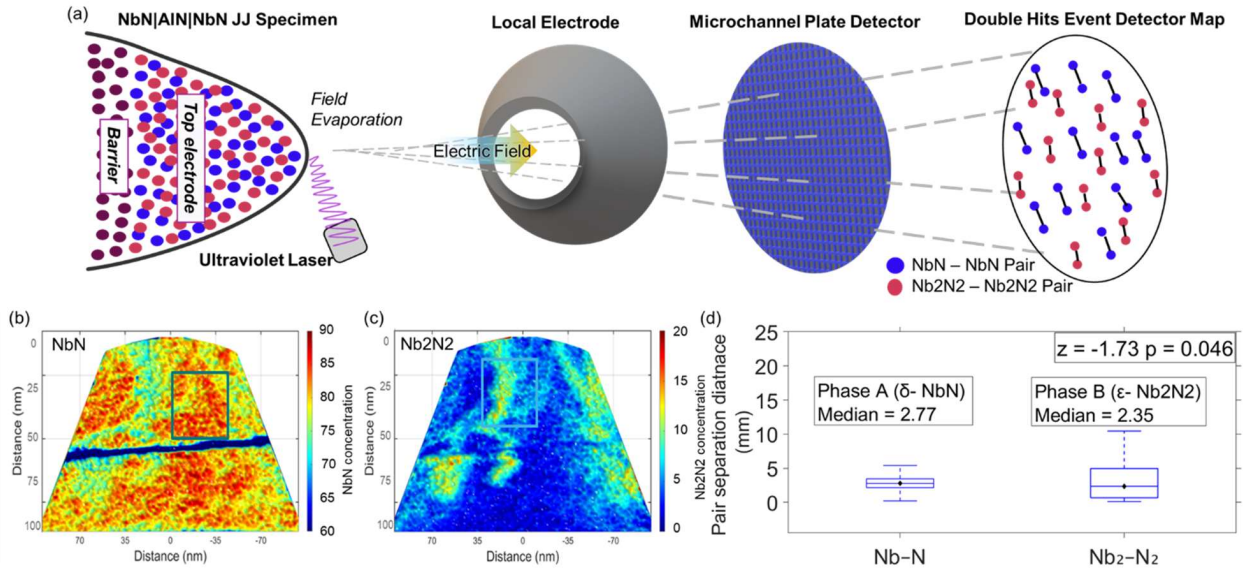

Figure 4: Workflow and results of APT based detector pair separation analysis of NbN/AlN/NbN trilayers. (a) Schematic of the APT double hit detection method, where simultaneously evaporated ions are projected onto the microchannel plate and analyzed as detector pair separations. (b) Spatial map highlighting δ-NbN distributions. (c) Spatial map highlighting ε-$Nb_2N_2$ distributions (mapped according to the respective Nb2N2 colorbar), showing localized environments. (d) Boxplots of detector pair separation distance calculations for δ-NbN and ε-$Nb_2N_2$, demonstrating relative bond length calculation for ε-$Nb_2N_2$ and δ-NbN.

The depth resolved Z-segmentation analysis (**Figure 5**) reveals the origin of these performance limiting defects, showing a systematic increase in the δ-NbN fraction from substrate to surface growth direction. Whereas ε-$Nb_2N_2$ fraction towards the substrate interface reaches a maximum of ~3.8% in the bottom electrode (90nm). This spatial localization strongly suggests its formation is governed by the initial ALD nucleation. We hypothesize this originates from a combination of factors. First, thermodynamic instability & interfacial strain, pure δ-NbN phase is thermodynamically unstable below 1050°C, well above our 400°C ALD growth temperature, where hexagonal phases like ε-$Nb_2N_2$ are thermodynamically preferred in addition to δ-NbN.[37,43] Furthermore, the significant lattice mismatch (~12.7%) between δ-NbN and sapphire likely induces strain, favoring nucleation of better matched hexagonal ε-polymorph (~7.3). Second, due to vacancy formation & impurity stabilization, the transformation pathway from cubic δ-NbN to hexagonal phases involves ordered nitrogen vacancies.[34] It is plausible that ALD chemistry may inherently create N vacancies at the interface, triggering this transformation, potentially aided by chemical stabilization from Al diffusion from sapphire and residual oxygen in the chamber. The low overall ε-$Nb_2N_2$ phase fraction (~4%) suggests strain relaxation, allowing the kinetically favored δ-NbN phase to dominate subsequent growth. This indicates a progressive destabilization of ε-$Nb_2N_2$ like local bonding environments during growth.

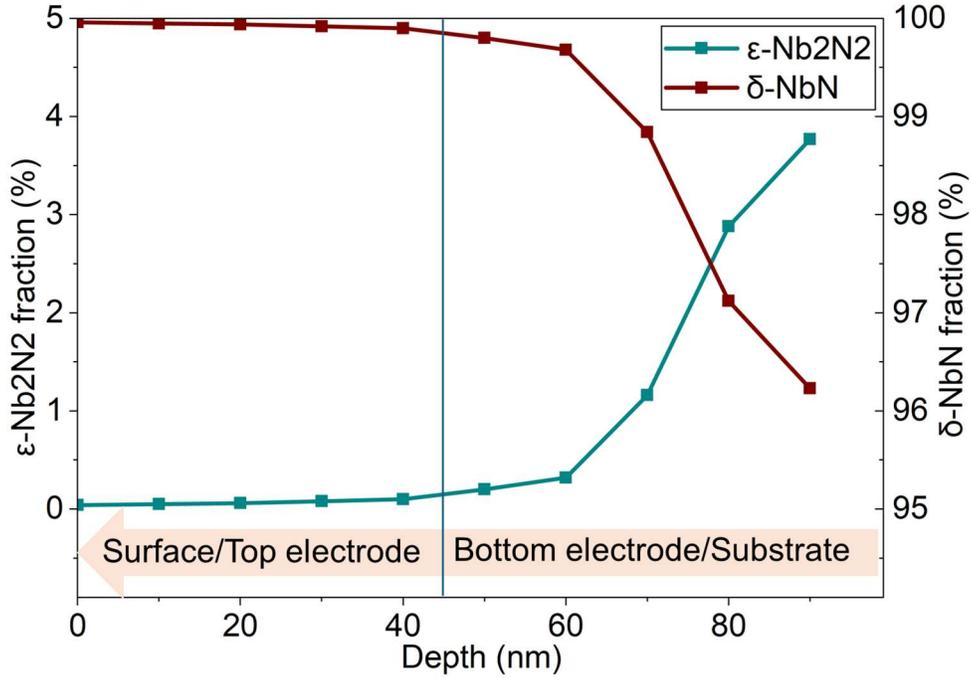

Figure 5: Depth-resolved fraction of ε-$Nb_2N_2$ and δ-NbN like bonding environments extracted from APT double hit analysis. A monotonic increase of δ-NbN from the substrate to the surface (along growth direction shown by the arrow) is indicated with a decrease of ε-$Nb_2N_2$.

This understanding points directly to potential mitigation pathways, targeted engineering of the critical substrate-film interface. Future work will focus on optimizing the ALD nucleation stage to suppress ε-$Nb_2N_2$ phase formation. Potential strategies could include the deposition of lattice-matched seed layer or tailored initial ALD cycles to prevent vacancy formation and promote the immediate phase pure δ-NbN stabilization. Controlling this ε- phase coexistence is critical to meet the benchmarks of high quality δ-NbN junctions, characterized by sharp gap features, enabling promising qubit coherence.[3,33] From a quantum information science perspective, the coexistence of these two phases with different $T_c$ values (~17 K for δ-NbN and ~11.6 K for ε-$Nb_2N_2$)[9,41], inherently creates spatial gap inhomogeneity. Such inhomogeneity can introduce additional electronic noise limiting qubit coherence times.[3,40] While previous ALD NbN studies focused on electrical transport or morphology alone, our combined APT and TEM analysis demonstrates phase competition is an intrinsic feature needing active control for high coherence, scalable devices. This work's significance lies in directly connecting atomistic phase coexistence with quantum device behavior, establishing a crucial feedback loop between chemistry, structure, and superconducting properties essential for fabricating reproducible JJs, and advancing quantum information sciences.

## Conclusion

In summary, we have established a transferable framework by integrating advanced materials characterization with macroscopic device analysis. This approach allowed us to identify a key microscopic origin of performance variability in ALD grown superconducting NbN/AlN/NbN JJs. Our combined APT and TEM analysis reveals the coexistence of nanoscale hexagonal $\varepsilon$-$Nb_2N_2$ polymorph inclusions within the desired $\delta$-NbN superconducting electrodes. The disparity between these two phases may account for the unresolved supercurrent and the soft subgap to normal state transition. Furthermore, the difference in superconducting properties between the $\delta$-NbN and $\varepsilon$-$Nb_2N_2$ can potentially create spatial gap inhomogeneity, introducing decoherence channels for the nitride qubits. This study not only identifies a specific, performance limiting defect but also traces its origin to the initial ALD nucleation at the substrate interface. The statistical methodology introduced here, including the bond length mapping, PCA based clustering, and Z segmentation analysis offers generalizable tools to probe spatial polymorphic coexistence in ultrathin superconducting electrodes. By directly linking atomistic phase coexistence to electronic behavior, this work provides a generic method to diagnose and mitigate structural defects in quantum devices. These insights lay the foundation for phase engineered NbN electrodes with enhanced coherence and reproducibility, providing a pathway toward scalable superconducting electronics and next-generation quantum information technologies.

## Methodology

### Atomic layer deposition

The ALD process was carried out using a Ultratech/Cambridge Fiji G2 plasma system in Yale University cleanroom. Sapphire substrates were loaded into the chamber after sequential cleaning with organic solvents and piranha. Prior to deposition, the substrates underwent $H_2/N_2$ plasma surface activation. The temperature of ALD chamber was set as 400°C. Tert-butylimido tris(diethylamido)niobium (TBTDEN) was the precursor for NbN deposition, while trimethylaluminum (TMA) served as the precursor for AlN. The plasma-enhanced deposition process utilized $H_2/N_2$ as reactive gases.

### Transmission electron microscopy

High resolution analytical (scanning) transmission electron microscopy (TEM) was carried out using an FEI (Thermo Fisher) TITAN (300 kV), equipped with energy-dispersive X-ray spectrometry (EDX). TEM sample preparation was performed using the lamella preparation method (lift-out) with a focused ion beam. A Helios 5 UC dual-beam electron microscope equipped with a manipulator and gas injection system was used for TEM sample preparation. A platinum protective coating was applied to prevent Ga contamination, and low-kV milling at 5 kV was performed in the finishing stages of the lamella. Further thinning of the lamella and reduction of gallium damage were carried out using a Fischione Nano Mill, equipped with Ar ion bombardment at different voltages (500–700 eV) for 20 min in each voltage.

### TEM data analysis for bond length mapping

TEM images were used to probe local structural variations across the NbN/AlN/NbN trilayer. Post-acquisition analysis was carried out in Digital Micrograph, where multiple 4.5 × 4.5 nm regions were randomly selected and processed by inverse fast Fourier transform (FFT) filtering to isolate lattice fringes and suppressing background noise. Corresponding live FFT patterns were converted into line profiles, capturing the periodicity and intensity oscillations associated with local d-spacings. To reduce noise, profiles were smoothed and their envelopes extracted, and both raw and envelope features were used for clustering analysis via k-means. Robust statistical filtering, including envelope percentile thresholds and physical range clipping, was applied to suppress artifacts. This envelope clustering approach grouped regions with similar structural signatures and enabled visualization of per-cluster boxplots, showing nanoscale variations in lattice periodicity.

### Atom probe tomography

Atom probe tomography (APT) was carried out using CAMECA LEAP 6000XS instrument in laser mode (laser wavelength ($\lambda$)=355nm) with a laser energy of 100 -120pJ, specimen temperature of 50K, pulse repetition rate of 250kHz, and 0.5% detection rate. The needle shaped APT specimens were prepared using Thermo Scientific Helios 5 plasma focused ion beam (FIB). The sharp APT needles were prepared by following standard site-specific lift-out and annular milling methods.[44] Multiple datasets were collected across the sample to ensure reproducibility, a representative reconstruction is shown in this work for reporting the structural and chemical quality of the layers.

APT data analysis for detector pair separation distance measurement

APT data analysis was conducted using CAMECA's Integrated Visualization and Analysis Software (IVAS 3.8). Analyzed data (EPOS) was imported in MATLAB using a custom MATLAB pipeline, and multiplicity values were recast to identify unique double hit events. Events were filtered within a defined region of interest (ROI) in detector and Z space. Following the approach introduced by Olivia et al.,[45] pairwise detector separations were calculated to obtain bond length distributions, and ion identities (NbN=R3, $Nb_2N_2$=R18) were assigned using mass spectrum range files. NbN – NbN (R3R3) and $Nb_2N_2$ – $Nb_2N_2$ (R18R18) double hits were isolated, and detector separation between these pairs were recorded. Kernel density estimates (KDE), and the resulting distributions were represented as boxplots with median bond lengths extracted for comparison. To test for statistical significance between these pairs, a non-parametric Mann – Whitney U test was applied, ensuring robustness against unequal sample sizes and non-normal distributions. The p-value reflects the probability of observing such a difference by random chance under the null hypothesis (no difference between groups), and a value below 0.05 indicates statistical significance. The z-value represents the standardized score of the observed difference relative to the expected distribution, with negative z here indicating that $\epsilon$-$Nb_2N_2$ separations are toward shorter distances compared to $\delta$-NbN. The Mann-Whitney U test is particularly suited to this analysis because the $\epsilon$-$Nb_2N_2$ population is relatively small and may not follow a normal distribution.[46] A feature matrix (bond length, stoichiometry flag, spatial midpoints) was constructed and analyzed using Principal Component Analysis (PCA) followed by DBSCAN clustering to reveal chemically distinct environments. Finally, bond statistics were evaluated as a function of Z-position to capture spatial variation in phase distribution opposite to the growth direction from top electrode towards bottom electrode.

### Electrical measurements

NbN/AlN/NbN Josephson junctions were fabricated by a flip-chip technique. Devices were wire-bonded to a printed circuit board for electrical measurements. Room-temperature resistances were measured, and several devices were selected for detailed current–voltage (I-V) characterization at 4 K. Devices were voltage-biased using a voltage source (QDAC-II). The current flowing through the devices was amplified by a current preamplifier (SR570) and subsequently recorded by a digital voltmeter (SIM970).

### Acknowledgement
We thank Karthick Gothandapani for performing STEM experiments. This work was supported by Air Force Office of Sponsored Research under Grant no. 162023-22608. D.W. and H.T. acknowledge the support from Air Force Office of Sponsored Research under Grant No. FA9550-23-1-0688, the Army Research Office under Grant No. W911NF-24-2-0240, Office Naval Research under Grant No. N00014-23-1-2021.

Table of Content

Atomistic multimodal analysis uncovers nanoscale ε-$Nb_2N_2$ inclusions and interfacial oxygen in ALD δ-NbN trilayers, revealing the structural defects that suppress superconducting transport and limit qubit coherence.

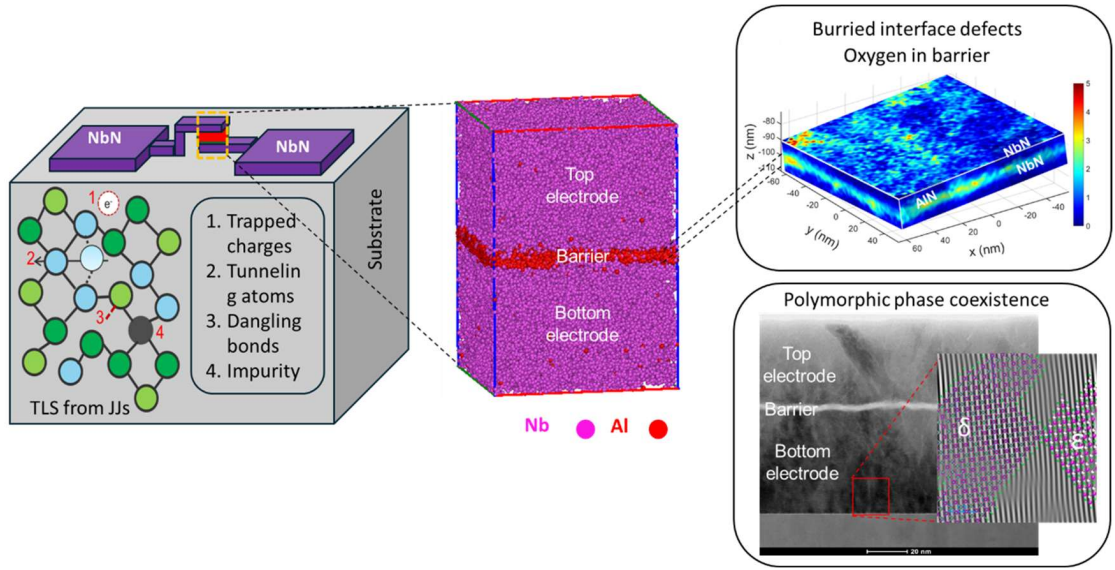